\documentstyle[aps,prb,epsf,graphicx,here]{revtex}
 
\begin{document}          


\title{Controlling the accuracy of the density matrix renormalization group method: The Dynamical Block State Selection approach}

\author{\"O.~Legeza,\footnote{permanent address: \\
Research Institute for Solid State Physics, H-1525 Budapest, P.\ O.\ Box 49, Hungary}
J. R{\"o}der and B.~A.~Hess}

\address{Chair of Theoretical Chemistry, Friedrich--Alexander University Erlangen--Nuremberg. D-91058 Erlangen, Egerlandstr. 3, Germany}

\date{\today} 

\maketitle

\vskip -8pt
\begin{abstract}
We have applied the momentum space version of the Density Matrix Renormalization 
Group method ($k$-DMRG) in quantum 
chemistry in order to study the accuracy of the algorithm in the new context. 
We have shown numerically that it is possible to determine the desired accuracy 
of the method in advance of the calculations by dynamically controlling the truncation 
error and the number of block states using a novel protocol which we dubbed
Dynamical Block State Selection (DBSS). The relationship between the real error and 
truncation error has been studied as a function of the number of orbitals and 
the fraction of filled orbitals. 
We have calculated the ground state of the molecules 
CH$_2$, H$_2$O, and F$_2$ as well as the first excited state of CH$_2$. 
Our largest calculations were carried out with 57 orbitals, the largest 
number of block states was 1500--2000, and the largest dimensions of the Hilbert 
space of the superblock configuration was 800.000--1.200.000.
\end{abstract}



\twocolumn

\section{Introduction}

Since its first appearance in 1992, the Density Matrix Renormalization Group method
\cite{white1,white2} has witnessed great developments and it soon became 
one of the most widely applied numerical methods in one-dimensional solid 
state physics. Within a short period of time, the real space renormalization 
method had been further extended and  
the momentum space version of the method ($k$-DMRG) was introduced by Xiang  
\cite{xiang} in 1996. Unfortunately, test calculations on the Hubbard model indicated
relatively poor performance compared to the real space version which hindered
further application of the method for several years. 

Quite recently, DMRG was used to study models of cyclic polyenes \cite{fano}  
and models of polyacetylene \cite{bendazzoli}.
S.~R.~White has successfully applied $k$-DMRG in quantum chemistry 
to calculate the ground state energy of molecules represented in the 
framework of the usual Linear Combination of Atomic Orbitals (LCAO) 
approximation, using small basis sets 
\cite{white3,white4}. His results seemed
challenging and attracted considerable attention which stimulated various 
groups \cite{mitrushenkov,chan} to start to work on the new field.   

Among all the various models studied by DMRG during the past decade the 
accuracy of the algorithm has always been a problem which is still not 
satisfactorily solved.  
The recent application of DMRG in quantum chemistry gives 
further grounds for benchmark investigations of this question within the new 
framework. 
In all attempts so far,  the accuracy of the method was analyzed a posteriori by
means of comparison with the corresponding full CI (FCI) benchmark results.  
For instance, recently, Chan {\em et al.} 
 \cite{chan} reexamined the scaling behavior of the real error, developing 
an extrapolation approach as a function of the number of block
states ($M$).

In this paper we show that in contrast to previous approaches, the 
desired accuracy of a DMRG calculation can be established in advance  
if we take into account the dynamic change of the 
reduced density matrix of the subsystem. Within our approach, described in the next section, 
we will be able to show that if the number of block states is adjusted dynamically,
a linear relationship obtains between the logarithm of the real error 
and the truncation error, which, in turn, can be used to derive a novel method 
to extrapolate to the full CI result.

Our main goal in this paper is to determine the accuracy of $k$-DMRG in 
quantum chemistry and show that the algorithm converges to the accuracy that 
was set up in advance of the calculation. We have, therefore, carried out a 
detailed DMRG study of CH$_2$, H$_2$O and F$_2$ molecules with various 
number of orbitals each representing different test cases. We have also addressed 
problems related to the initial block-state configuration
that arise within the framework of $k$-DMRG. Since the focus of the paper 
is on the dynamic scaling of the density matrix and parameters of DMRG, we 
recall only those main definitions and formulas in this paper that are relevant
to the question and not well known in quantum chemistry. Therefore, details of 
our numerical procedure and developments will 
be published elsewhere. Although we have analyzed the general trend of the 
numerical error of $k$-DMRG through quantum-chemical calculations, our results can be 
generally applied to other quantum system as well.

The setup of the paper is as follows. In Sec.~II we briefly describe the main 
steps of DMRG and recall the main sources of the numerical error. Sec.~III 
is devoted to the details of the numerical procedure used to determine the dynamic 
scaling behavior of the density matrix and to the problems that appear in the context
of quantum chemistry. Sec.~IV contains the numerical results and analysis of the 
observed trends of the numerical error. The summary of our conclusions 
and a few general comments about the algorithm is presented in Sec.~V.

\section{Background of the numerical error}

Detailed description of the DMRG algorithm can be found in the original papers
\cite{white1,white2,xiang} and its application in the context of quantum 
chemistry is summarized in two recently published papers \cite{mitrushenkov,chan}. 
Therefore, we present only the most important formulas and definitions that are relevant to 
the question of accuracy. 

The main purpose of DMRG is to treat the electron--electron correlation in a rigorous  
way which allows the minimization of the energy and calculation of 
measurable quantities. Since DMRG is a variational procedure, it always provides 
an upper bound for all the calculated quantities.  
In the context of quantum chemistry a one dimensional chain that is studied by DMRG 
is built up from the
molecular orbitals that were obtained, e.g., in a Hartree--Fock calculation. 
The electron--electron correlation
is taken into account by an iterative procedure that minimizes the Rayleigh quotient 
corresponding to the Hamiltonian describing the electronic structure of the molecule, 
given by 
\begin{equation}
{\cal H} = \sum_{ij\sigma} T_{ij} c^\dagger_{i\sigma}c_{j\sigma} + 
           \sum_{ijkl\sigma\sigma^\prime} {V_{ijkl} 
           c^\dagger_{i\sigma}c^\dagger_{j\sigma^{\prime}}c_{k\sigma^{\prime}}c_{l\sigma}}
\label{eq:ham}
\end{equation}
and thus determines the full CI wavefunction. 
In Eq.~(\ref{eq:ham}) $T_{ij}$ denotes the matrix elements of the one-particle Hamiltonian 
comprising kinetic energy and the external electric field of the nuclei,  
and $V_{ijkl}$ stands for the matrix elements of the electron repulsion operator.
In order to show what are the key concepts and parameters of the numerical renormalization
procedure and what are those drawbacks which hinder the analytical study of the method 
we have included a brief overview of the renormalization group methods.

\subsection{Block renormalization group method (BRG)}

In order to determine the eigenvalue spectrum of the Hamiltonian 
corresponding to an infinite long quantum 
chain (in the context of quantum chemistry this means infinitely many orbitals) 
built up from quantum sites represented by $q$ basis states,
blocks were formed from each of two adjacent sites, and the Hamiltonian was determined
on the new configuration as is shown on Fig.\ \ref{fig:ren}. 
First the Hamiltonian of the model is diagonalized for two sites and 
then the $q$ lowest energy states are selected out of the $q^2$ states whereby the so 
called block site will represent the two-sites problem in the subsequent 
iteration step. Operators defined on the selected $q$ basis states are obtained
from the original site operators according to a renormalization procedure 
given by the equation 
\begin{equation}
A_{ren} = O A O^\dagger,
\label{eq:renorm}
\end{equation}
where operator $O$ is constructed from the selected $q$ eigenfunctions of the 
two-sites problem. In order to retain the original structure of the Hamiltonian 
operator, on-site (in the figure labeled by $h$) and inter-site (denoted by $\lambda$) 
coupling constants are renormalized as well, shown as $h^\prime$ and $\lambda^\prime$. 
In the subsequent step 
the $q^2$-dimensional Hamiltonian operator is diagonalized for two adjacent 
block sites, and again $q$ states with lowest energy are 
selected out for the block site that will represent four sites in the 
following iteration step. Since the structure of the original Hamiltonian 
operator is retained and the number of coupling constants is unchanged,
changes of the coupling constants (flow 
equations) can be studied analytically. When subsequent iteration of the 
renormalization steps leaves the coupling constants unchanged, the algorithm 
has reached a fix point which represents the infinite length (thermodynamic) 
limit of the model.

\subsection{Wilson's renormalization group method}

\begin{figure}
\includegraphics[scale=0.35]{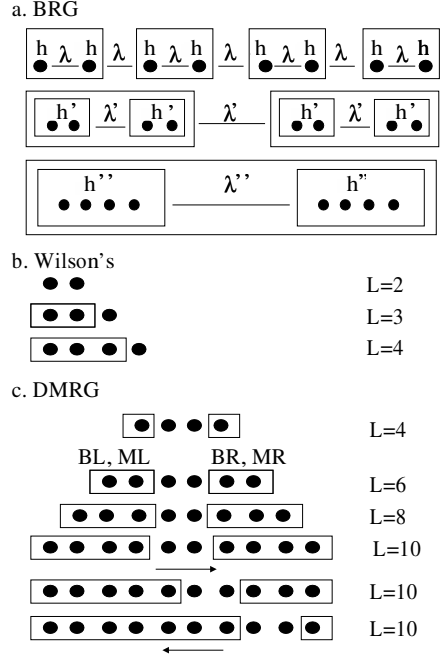}
\caption{Schematic plot of the spin couplings in the BRG, Wilson's and DMRG
renormalization methods.}  
\label{fig:ren}
\end{figure}

Besides a few analytically solvable models it turned out that the BRG method can
be used only numerically and its systematically increasing inaccuracy 
hindered the application of the method.     
In 1975 Wilson introduced another procedure for the numerical renormalization
method \cite{wilson1,wilson2}, in which a quantum chain with finite length $L$ is built up 
systematically from quantum sites represented by $q$ basis states 
by keeping the size of the Hilbert space fixed as is shown on 
Fig.\ \ref{fig:ren}. 
The main idea of 
the method was again to solve the Hamiltonian of the model for two sites and
selecting $q^\prime$ lowest energy states out of the $q^2$ states where $q^\prime$
was increased systematically up to a maximum value during the first few iteration steps 
based on the energy spectrum and kept constant afterwards.  
Operators were renormalized according to Eq.\ (\ref{eq:renorm}). The key difference of Wilson's
method compared to BRG is that he did not retain the original structure of the 
Hamiltonian operator, but he analyzed the scaling behavior of the energy as a
function of the chain length. Systematical application of the renormalization
procedure introduces new terms and coupling constants. However, many of them
become irrelevant for longer chains and the method also drives the system into
the fixpoint. The major drawback of the method is that since the structure of
the Hamiltonian changes with increasing chain lengths, flow equations
can not be defined and the method can not be studied analytically.   
 
\subsection{Density matrix renormalization group method (DMRG)}

In spite of the powerful properties of Wilson's procedure, 
the numerical error of the method 
grew systematically with increasing chain length, which drawback has led to 
the fact that longer chains could not be studied numerically. Besides the 
truncation of the Hilbert space through the renormalization procedure, the 
numerical error had another main source. When an additional unrenormalized 
site was added to the 
block site, the coupling was taken into account only between the block site 
and this new site. In each iteration step the problem was, therefore, reduced 
to an isolated two-sites problem with open boundary conditions. These 
observations has led S.~R.~White to construct a larger auxiliary system 
(superblock configuration) which contains an environment in addition to 
the original block site problem to take care of the boundary effects in a 
more reliable way, as is shown on Fig.\ \ref{fig:ren}. 
According to the figure the structure of the superblock configuration is defined 
as $B_L \bullet \bullet B_R$
where $B_L$ represents the block site, $\bullet$ the new site under consideration, 
the additional $\bullet B_R$ configuration the environment and $M_L$ and $M_R$
denotes the number of block states, respectively.  
In order to minimize the error introduced in the representation of the block 
state in the truncation process, S.~R.~White has constructed the $O$ matrix 
using the eigenfunctions of the reduced density matrix of the subsystem $B_L\bullet$.
It has been recognized in different context \cite{hess1} that the reduced subsystem 
density matrix describes the interactions of two subsystems in a particularly 
efficient way. Using these  
two key ingredients, a DMRG iteration step first includes the diagonalization
of the Hamiltonian 
constructed on the superblock configuration to obtain the target state.
The target state is chosen from the eigenvalue 
spectrum of the Hamiltonian that we want to calculate. It can also be a linear 
combination or even an incoherent superposition of more eigenstates as well. 
If $|I\rangle$ and $|J\rangle$ denote basis states for
$B_L \bullet$ and $\bullet B_R$, respectively, then the target state is
written as
\begin{equation}
\Psi_{Target} = \sum_{I,J}^{M_L*q,M_R*q} \psi_{I,J} |IJ \rangle,
\label{eq:psi}
\end{equation}
where $\psi_{I, J}$ is determined by diagonalization of the superblock 
Hamiltonian. 
After the target state is obtained, 
the reduced density matrix of the $B_L \bullet$ subsystem 
\begin{equation}
\rho_{I,I^{\prime}} = \sum_J \psi_{I,J}\psi_{I^{\prime},J}
\label{eq:rho}
\end{equation}
is diagonalized and the $M$ eigenstates with largest eigenvalues 
($\omega_{\alpha}$)
are selected to build up the $O$ matrix. The site operators are 
renormalized according to Eq.~(\ref{eq:renorm}). 
The error of the truncation procedure in the DMRG method can be measured 
by means of the deviation of 
the total weight of the selected states from unity which is defined as
\begin{equation}
TRE = 1 - \sum_{\alpha=1}^{M_L}\omega_{\alpha}.
\label{eq:tre}
\end{equation}

The initial $B_L$ and $B_R$ configuration contain one site per block each, thus the 
superblock Hamiltonian is determined on $q^4$ basis states restricted to 
the conserved quantum numbers like the total spin or the number of electrons. 
In each iteration step the size of the chain is increased by two sites until 
the desired chain length is reached as is shown on Fig.\ \ref{fig:ren}. 
This procedure is the so called {\em infinite lattice} algorithm. 
In order to average out long-wavelength fluctuations, the 
superblock configuration is asymmetrised by increasing the size of $B_L$
and decreasing the size of $B_R$ until the left block contains 
$L-3$ sites and the right block one site. The same procedure is then carried 
out in the reverse way and when the configuration is symmetric again, the first 
sweep of the so called {\em finite lattice} algorithm ends. This procedure can be 
repeated infinitely many times and is usually stopped when the energy does not change
within two subsequent sweeps. 
There is again a major difference between BRG and DMRG which makes the analytical 
study of the scaling behavior of the latter method very complicated: In the 
DMRG method the number of selected block states ($M$) is larger then $q$ 
and the original structure of the Hamiltonian is not retained, thus flow 
equations of the coupling constants can not be determined.

According to the two key ingredients of the method, the numerical error of 
the DMRG algorithm has basically two independent components which are the 
truncation error and the environmental error. The first one is generated during
the renormalization step due to the truncation of Hilbert space, while the 
environmental error appears because the chain is built up from blocks and the 
long range interactions are cut off. As it was shown in Ref.~\onlinecite{legeza1} using
the {\em finite lattice} method, the environmental error can be averaged out 
and 
finally there remains a linear relationship on a log--log scale between the 
real error and truncation error.

The truncation error, on the other hand, strongly depends on the shape of the 
eigenvalue spectrum of the reduced subsystem density matrix and on the number of block 
states kept for the subsequent iteration step. It has also long been known 
that the structure of the density matrix depends on the criticality of the model. 
For systems with finite energy gap and coherence length the density matrix 
eigenvalue spectra decays exponentially, while for critical models with 
infinite coherence length it has a power-law tail. Besides these, in case of 
analytically solvable models the structure of the eigenvalue spectra of the 
density matrix determines the energy spectrum of the model as it was shown 
in Ref.~\onlinecite{kaulke}.

In addition to all the points discussed above, the decay of the eigenvalue spectrum 
also changes as the target state gets closer to the exact solution. It is, 
therefore, evident that selecting out the $M$ most probable states with 
highest eigenvalues will be an insufficient condition to control the accuracy 
of the DMRG method. Instead, one has to take care of the dynamic changes of 
the spectrum of the density matrix and keep the truncation error below a given
threshold.  
Since the structure of the density matrix represents the whole system as well, 
it naturally arises that the number of block states should be selected out in 
a way that the truncation error satisfies an initial condition that 
was introduced in advance of the calculation. 

\subsection{QC-DMRG method}

In the context of quantum chemistry, a one dimensional chain containing $L$ 
molecular orbitals is generated by ordering the orbitals employed to build 
up the multi-particle states with 
increasing energy or by other rules, analogous to $k$ points in $k$-DMRG \cite{xiang}.
These molecular orbitals
are calculated by standard numerical methods of quantum chemistry. 

It worth to note that the optimal ordering of the orbitals in the chain 
is still an open field of research. Note that the initial 
chain length of the QC-DMRG is $L$ from the very beginning and the block operators 
for the left and right blocks  
are generated by a ``warm up'' procedure \cite{xiang}  
instead of the {\em infinite lattice} algorithm. The effect of the electron--electron
correlation is taken into account by the systematic sweeps in the framework of the 
{\em finite lattice} algorithm. 
Since the overall performance of the QC-DMRG method differs from the real-space version, 
it is also expected that 
new problems arise due to the inaccuracy of the starting wave function. These 
will be also investigated in detail in the next section.

The most straightforward procedure to represent the unrenormalized site operators 
is to define them on spin-orbital basis states, in which case $q$ is equal to two.
The phase operator is then taken care of automatically by the standard definition
of fermion creating and annihilating operators. On the other hand, 
if orbitals from, e.g., a restricted Hartree--Fock (RHF) calculations are employed, 
it is possible to define a super site built up from the ordered tensor product
of spin-down and spin-up basis states, in which case $q$ is 4 and the phase 
factor must be explicitly taken care of. This method offers considerable efficiency gains
because in this way the chain is only half the size compared to an unrestricted HF (UHF)
type formulation, using spin orbitals for each site. Thus 
the number of multiplications using quadratic auxiliary operators during the 
superblock Hamiltonian diagonalization procedure \cite{white3} is roughly reduced 
by a factor of 4 compared to the spin-orbital formulation.  
In our implementation we have built up the chain from super sites.

\section{Controlling the absolute error of DMRG}

\subsection{Dynamic adjustment of the number of block states}

In order to control the accuracy of the DMRG procedure, the selection of the 
multi-particle states of the superblock Hamiltonian which are used for renormalization 
is obviously the decisive issue. 
Keeping all states featuring eigenvalues of the subsystem reduced density matrix
larger than a fixed parameter which we called 
$DM_{cut}$ during the renormalization procedure, the truncation 
error can be as small as $DM_{cut}$, but it can be larger if the integrated 
contribution of the neglected states is still significant.  
To avoid such problem we propose to adjust $DM_{cut}$  
dynamically, thus the number of selected states is increased
as long as the integrated weight of neglected states is larger then a maximum value
$TRE_{\rm max}$, which can be fixed at the beginning of the
calculation. This enables us to set up the desired accuracy of the DMRG algorithm 
at the beginning of the calculation. The  
number of states will be adjusted by this protocol in a dynamical fashion,
depending on the structure of the density matrix spectrum.

Since the truncation error is not immediately connected to the error in energy,
one can control only the relative error in this way. In 
order to control the absolute error in energy, $TRE_{\rm max}$ should be
scaled by the Hartree--Fock energy or by the energy value calculated by  
the DMRG method which usually has the same order of 
magnitude as the exact value even after the first few iterations.
We then expect the relative error of the energy to converge to this scaled 
threshold within a few sweeps of the DMRG procedure.

From technical point of view, dynamic selection of block states has another 
important advantage. 
In the standard DMRG calculation the number of block states is fixed. Using
our dynamical adjustment, the largest number of block states required to 
guarantee a given truncation error develops, however, only close to the 
symmetric configuration during the sweep. For most of the remaining steps
the threshold $TRE_{\rm max}$ is reached with considerably smaller number of 
block states, leading to substantial gains in efficiency in the
renormalization step and the construction of the next superblock Hamiltonian,
when dynamic block state selection is used.

Within the framework of our procedure, it is also evident 
why previously developed extrapolation methods based on functions of the number 
of block states failed to estimate the scaling behavior of the error in a 
rigorous way. The value of $M$ is only one of the factors that determines the 
largest value of the truncation error during a full sweep. Using it exclusively,   
changes of the density matrix are not taken care of. Thus it is almost 
impossible to derive a reliable formula to estimate the real error as a 
function of the number of block states for the general case.

\subsection{Initial condition for the number of block states}

The straightforward application of dynamical control of $DM_{cut}$ during the first 
few sweeps is complicated 
by the fact that there is a major difference between the wave 
function of a given chain length generated by the {\em infinite lattice} algorithm 
of the real space version and that of generated by ordering the orbitals in 
the case of $k$-DMRG. In the first method, the wave function of the target 
state is always very close to the one which is obtained after several sweeps of the 
{\em finite lattice} method; however this is not true in general for the momentum 
space version when the wave function strongly depends on the ordering 
of the orbitals. For example, it typically happens that during the first few steps the 
density matrix eigenvalue spectra will have very few states with large 
eigenvalues and many states with almost zero weight. In this case, the number 
of selected states will be cut drastically, which will limit seriously the 
size of the Hilbert space in the subsequent iterations, causing the 
algorithm being trapped in a local minimum. This situation 
happens in other optimization methods as well, and for example in the case of 
simulated annealing the so called adiabatic heating is used to move the  
algorithm out from the attractor of a local minimum. 
In the context of DMRG the introduction of  
virtual states is required in this situation, which means that 
we keep also those states that had almost 
zero eigenvalue up to a fixed number that we called $M_{\rm min}$ during 
the first two sweeps. Usually after the first sweep the decay of the 
density matrix spectrum becomes smooth and it changes dynamically as the 
target state gets closer to the exact one.

\subsection{New criteria for convergence and extrapolation of the FCI energy}

Up to now the condition for the number of sweeps was determined in an 
empirical way, using the condition that the algorithm is stopped when 
the energy value obtained by two subsequent sweep does not change any more. 
Within the framework of DBSS 
we have a new criterion for the convergence. We 
have found that after convergence not only the energy value remains stable,  
but also the eigenvalue 
spectrum of the density matrix and thus the block states selected out by the 
algorithm for a given $B_L \bullet \bullet B_R$ configuration are the same 
during all subsequent sweeps. Although all subsequent sweeps leave the density 
matrix unchanged, still a fix point is not obtained since the structure of the 
density matrix and thus the truncation error and the relative error after 
convergence can slightly change (but within the same order of magnitude) 
depending on the initial condition, for example on different $M_{\rm min}$. On the 
other hand, we can treat the energy values obtained for various $TRE_{\rm max}$ 
values as points on a flow equation that converge to the fix point at the end, 
which is the FCI energy. Based on our previous results \cite{legeza1} and 
those presented in the next section we can extrapolate to the FCI energy using
the equation
\begin{equation}
\log \frac{E - E_{FCI}}{E_{FCI}} = a*\log (TRE) + b,
\label{eq:fcifit}
\end{equation}
where $a$, $b$, $E_{FCI}$ are parameters determined from the fit of the 
numerical result. As discussed below, our numerical results show that the
value of $a$ is close to one.

\subsection{Error of the excited states due to the inaccuracy of starting 
block states}

Besides the problem of the initial structure of the density matrix there is 
another difficulty which stems from the inaccuracy of the starting block wave 
functions. By contrast to the {\em infinite lattice} method when the target state 
always remains in the same spin symmetry or changes sign periodically as a 
function of the chain length \cite{legeza2}, the symmetry of the target state 
depends on the initial ordering in the case of $k$-DMRG. This can lead to a 
major error, because the DMRG algorithm can lose the target state if its 
symmetry changes during the first few sweeps. It can happen that for example 
targeting the second level the coefficients of the wave function of the ground 
state and excited states will mix and the spin symmetry of the target state 
changes randomly, and thus the energetically lowest level will be lost and the 
third level will become the target state.

\subsection{Introduction of local symmetry operators}

In order to avoid the random change of the spin symmetry we have introduced
partial spin adaption making sure that the permutational symmetry of the spins is odd 
for even $S$ and even for odd $S$, which implements 
the spin reversal operator that flips the spins along the $z$-directions
as it was shown in Ref.\onlinecite{legeza2}. In case of $k$-DMRG the starting 
block wave function is constructed in a way that it contains basis states 
with $N_{up}$ and $N_{down}$ quantum numbers, thus fixing $m_s$, 
and their symmetric components (i.~e.~, states with $-m_s$)
as well. During the renormalization procedure a state and its partner 
belong to same eigenvalue of the density matrix, thus the dynamic selection 
rule automatically ensures that both of them are kept. It worth to note, 
that this is not the full adaption of $S^2$ symmetry, which would be clearly
be desirable, but more complicated to achieve in the framework of DMRG. Thus,
components of the  singlet and quintets levels can still mix, but it is 
not a problem since they are usually well separated. Application the corresponding  
spin reversal operator effectively ensures that the target state will remain in the 
spin symmetry sector that was fixed at the beginning of the calculation. 

From technical point of view, this has the additional advantage that one needs to target 
only the first level in both spin symmetry sectors, which always requires less 
block states to achieve a given accuracy. In addition, the number of auxiliary 
operators needed during the diagonalization of the superblock Hamiltonian is 
decreased by a factor of two which doubles the speed of the algorithm. For the 
half-filled case the particle--hole symmetry operator can be introduced in the 
same way. Details of the numerical procedure will be published elsewhere.

\subsection{Error of the expectation value of one- and two particle operators} 

The expectation value of the one- and two electron operators can be 
calculated from the one particle density matrix according to  
\begin{equation} 
\langle A \rangle = TR(\rho A).
\label{expectation-values}
\end{equation}
where $A$ is a $L$ by $L$ matrix of operator for a first-order property (e.g.,
dipole moment) in the same representation
as the original $T_{ij}$ and $V_{ijkl}$ were. So with $A = T_{ij}$
Eq.~\ref{expectation-values} provides the kinetic energy of the FCI wave
function.
Once the target state was obtained, the one particle reduced density 
matrix can be formed for any  $B_L \bullet$ configuration as 
\begin{equation} 
\rho_{ij} = \langle \Psi_{Target} | \sum_\sigma  c_{i\sigma}^\dagger
c_{j\sigma} | \Psi_{Target} \rangle 
\end{equation}
where $i$, $j$ denote sites in the left block.  The one-particle density matrix 
for the right block is determined in a similar way. If $i$ is in the left block
and $j$ in the right block, then $\rho_{ij}$ is constructed from the one- 
particle operators of the two blocks. 
The latter case was used to calculate 
two-point correlation functions in real space DMRG and was shown that the
error of the one- and two-point correlation function is larger by one or two orders 
of magnitude compared to the error of the ground state energy. Since the dynamic block state 
selection rule controls the accuracy of the ground state, it also ensures the 
same scaling behavior of the correlation functions as well. Besides that, the 
fluctuation of the error shown in Ref.~\onlinecite{legeza1} because of the 
fluctuation of the truncation error within a full sweep caused by to the constant
value of $M$ also diminishes.   

The two-particle reduced density matrix can be obtained in a similar way
\begin{equation} 
\Gamma_{ijkl} = \langle \Psi_{Target} | \sum_{\sigma \sigma^\prime}
c_{i\sigma}^\dagger c_{j\sigma^\prime}^\dagger
c_{k\sigma^\prime} c_{kl\sigma}  | \Psi_{Target} \rangle,
\end{equation}
where the four-operator term is decomposed into four independent terms depending 
on the distribution of the $i,j,k,l$ indices along the chain making use of the
usual partially contracted operators of $k$-DMRG \cite{xiang}.

\section{Numerical results}

In order to study the performance of $k$-DMRG in quantum chemistry, 
we followed a route similar to the one we
used to study the accuracy of the 
real-space DMRG \cite{legeza1}. We  
performed calculations on molecules with different 
properties for which DMRG is expected to possess different scaling 
behavior. 
Thus we have carried out a detailed DMRG study of the absolute error of the
energy as a function of the number of orbitals and the fraction of filled 
orbitals on molecules CH$_2$, H$_2$O, and F$_2$.
The Hartree--Fock orbitals in a given basis of Gaussian orbitals were calculated,
and the $T_{ij}$ and $V_{ijkl}$ matrix elements were transformed to the 
Hartree--Fock basis using the MOLPRO program package\cite{MOLPRO}, which was also
used for the calculation of the benchmark full-CI energies \cite{FCI1,FCI2}.

We used various basis sets and geometries for the molecules which we
selected for benchmark calculations. The geometries, references to the
basis sets employed, and results obtained in SCF calculations as well
as full CI energies are detailed in Table~\ref{molecules}. The models
employed for the water molecule have also been used 
in White's study \cite{white3}. We include
these cases here in order to enable a direct comparison with previous work.
A more interesting test case was to study the CH$_2$ molecule, for which
we report energies for the triplet ground state as well as for the
first excited (singlet) state. Hartree--Fock orbitals of the closed-shell
singlet configuration were employed in all calculations on CH$_2$.
Calculations of the FCI energy of the triplet state were carried out
in both the $m_s=0$ and $m_s=\pm 1$ spin sectors. 
In order to show 
that the relative error scales to $TRE_{\rm max}$ independently
of the fraction of 
filled orbitals we have studied the half--filled chains by  
calculating the ground state of F$_2$ with 14 electrons and 14 orbitals 
(freezing the fluorine $1s$ orbitals and discarding the two highest
virtual orbitals) and with
18 electrons and 18 orbitals.
The latter calculation provides evidence that QC-DMRG is capable to
provide cutting-edge CASSCF calculations with the potential to push
their limits to active spaces well beyond a size which is feasible
nowadays by standard methods.

\subsection{Dynamic selection of Block states}

\begin{figure}
\includegraphics[scale=0.5]{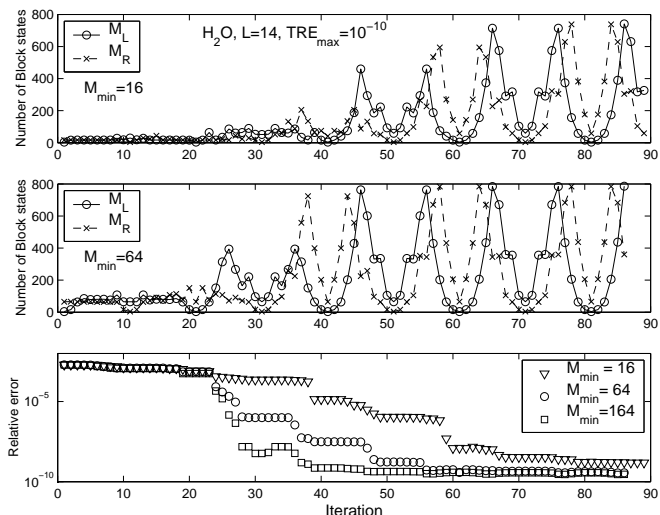}
\caption{The figure shows the dynamically selected number of left- and right 
block states $B_L$, $B_R$, respectively, for two values of the minimum threshold
value $M_{\rm min}=16, 64$ and the relative error 
as a function of iteration obtained with $M_{\rm min}=16,64,164$. In all cases the 
10 electrons of the H$_2$O molecule were correlated in the double-zeta water model
with 14 orbitals, and $TRE_{\rm max}=10^{-10}$ was set in advance of the calculations.}  
\label{fig:h2o_14}
\end{figure}

QC-DMRG calculations on the water molecule demonstrate the
dynamic selection of block states. In the first two panels of
Fig.~\ref{fig:h2o_14} we have plotted the number of block states
which were selected in a calculation correlating
10 electrons in 14 orbitals by means of QC-DMRG, starting with different
values of $M_{\rm min}$;

The number of block states for the left and right blocks is denoted by
$M_L$ and $M_R$, respectively. In the third panel of
Fig.~\ref{fig:h2o_14} we give the relative error 
($(E_{DMRG}-E_{FCI})/E_{FCI}$) of the calculation as
a function of $M_{\rm min}$ and the
the iteration step. The value of  
$TRE_{\rm max}$ was set to $10^{-10}$ in advance of the calculations. 
It is evident from the figure that the maximum number of
block state does not depend on the prescribed minimum value ($M_{\rm min}$), 
although it is reached faster for larger $M_{\rm min}$. 
In order to show that the converged value of the accuracy does not depend on the 
threshold value (once a large enough value was taken) we have also included the result obtained
with $M_{\rm min}=164$. It can be seen in the figure that the relative error
converges to $TRE_{\rm max}$ in all cases, 
but the speed of convergence strongly depends on $M_{\rm min}$. 
In order to show that the QC-DMRG algorithm is trapped in a local
minimum if $M_{\rm min}$ is chosen too small, we carried out calculations
with $M_{\rm min}=4,8$ and found indeed the number of block states being hindered to grow up.
Similar test calculations on longer chains indicated that a larger value of 
$M_{\rm min}=64$--$100$ is needed, thus  
we suggest that in order to 
to avoid problems related to local attractors and to obtain a 
faster performance a value of $M_{\rm min}$ no less than $150$--$200$ should 
be taken for longer chains.

Investigating the scaling of the relative error shown in the third panel 
of Fig.~\ref{fig:h2o_14}
one can find long plateaus where the accuracy of the method is not improved. In the
usual DMRG calculations going through such plateaus costs almost the same amount 
of time as calculating the region where the error drops significantly. 
By contrast, 
it can be seen on the figure that the minimum value
of $M$ occurs in the region of the
plateaus resulting in a very fast transversal of these regions.   
In addition, the maximum values of $M_L$ and $M_R$ occur at different iteration steps,
thus for a given superblock configuration we find that
even if one of them is very large, the other is usually much smaller. 
These two facts, finally, optimize the computational time and memory
resources within
a full sweep of the method.

\begin{figure}
\includegraphics[scale=0.5]{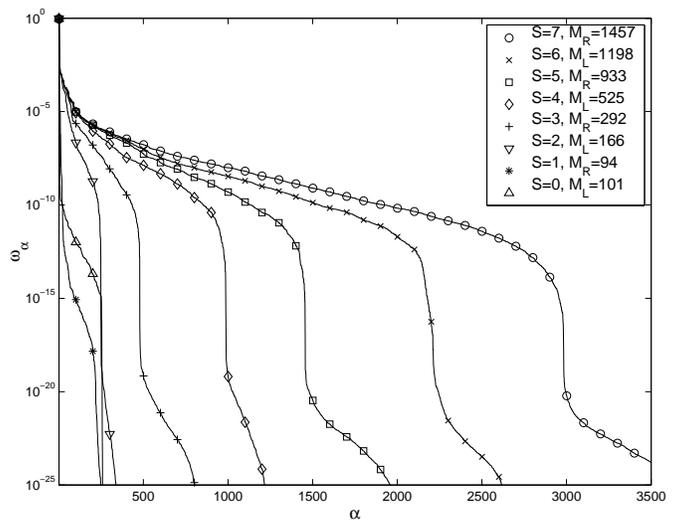}
\caption{The figure shows the eigenvalue spectrum 
of the reduced subsystem 
density matrix obtained for the F$_2$ molecule after the end of the sweeps($S$)  
of the {\em finite 
lattice} method. In the legend we have also included the number of 
selected block states ($M_L,M_R$) as a function of sweeps.}
\label{fig:f2_18}
\end{figure}

In order to show the dynamic change of the structure of the reduced subsystem 
density matrix, we have plotted in Fig.~\ref{fig:f2_18} 
the eigenvalues of the reduced subsystem density matrix obtained at the 
symmetric configuration (left and right blocks contained 6 orbitals) from a 
calculation of the F$_2$ molecule represented by 14 electrons and 14 orbitals.
There are several conclusions that one can draw from the figure. First of all,  
the density matrix spectrum decays very rapidly during the first few sweeps 
(S=0 is part of the ``warm up'' procedure ) which clearly implies the requirement of  
introduction of virtual states. On the other hand,  
as the target state gets closer to the FCI limit, the fraction of eigenvalues larger then 
$10^{-15}$ increases significantly. 
It can be seen from the figure that the decay of the spectrum can be fitted by  a linear line on a 
semilogarithmic scale for the largest eigenvalues, thus the 
density matrix spectrum decays exponentially, where the slope
is related to the finite coherence length of the model. On the other hand, 
the slope of the line changes as a function of sweeps until the algorithm converges. Once the relative
error converged to $TRE_{\rm max}$ (which means for $S>7$ in the case at hand) 
the slope of the decay remains the same, and this is the reason why the 
number of selected block states are the same for the subsequent sweeps. 
It worth to note that since the decay of the density matrix can be fitted 
by a straight line in this
model, the truncation error can be estimated as a function of the block states. However, in order
to obtain a rigorous scaling behavior of the error as a function of block states one has to
include the change of the slope as well,
which in general strongly depends on the static and dynamic correlations of the models.

\subsection{Relationship between the relative error and $TRE_{\rm max}$}

\begin{figure}
\includegraphics[scale=0.5]{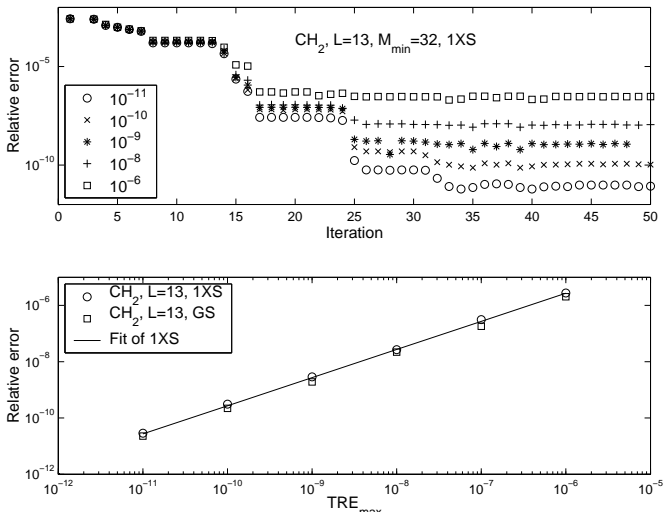}
\caption{Calculations for CH$_2$ with $L=13$ sites shows the relationship between 
relative error and $TRE_{\rm max}$. The straight line is the result of the fit.
GS denotes the triplet gound state, and 1XS denotes the first excited (singlet)
ground state.
} 
\label{fig:ch2_13}
\end{figure}

In order to test that the relative error converges to a given value of $TRE_{\rm max}$
we have ran independent calculations for all the test molecules by adjusting 
$TRE_{\rm max}$ from $10^{-3}$ up to $10^{-11}$. The relative error of the first excited 
state 
obtained for the CH$_2$ molecule with 6 electrons and 13 orbitals using $M_{\rm min}=32$
as a function of the iteration step and $TRE_{\rm max}$ is shown on 
Figure~\ref{fig:ch2_13}.  
It can be seen on Fig.~\ref{fig:ch2_13}a. 
that the relative error of the first excited state also 
converges to the values of $TRE_{\rm max}$ set up in advance of the calculations.
The converged value of the relative error as function of $TRE_{\rm max}$ 
for the first excited as well as for the ground state
is plotted on Fig.~\ref{fig:ch2_13}b.  
It is clear from the figure 
that there is a linear relationship between the converged value of the relative error 
and the truncation error, the fitted slope being 0.98. 
Fitting our results obtained for the various tests cases also with different 
$M_{\rm min}$ values, we have found that the slope was always between $0.95$ and $1.1$. 
Calculations in the $m_s=\pm 1$ spin sectors provided a faster convergence for the
ground state, as expected. The residual splitting of the $m_s=0,1,-1$ components of the
triplet level was as low as $10^{-12}$ a.u. 

Calculations performed on the other test molecules with different number of basis 
states and for various values of $M_{\rm min}$ showed that the relative 
error scales to $TRE_{\rm max}$ independently of the 
number of orbitals, fraction of filled orbitals and the threshold level of the number
of block states.  
Of course, the convergence gets slower for longer chain lengths 
and we usually needed 6--8 sweeps to gain an absolute accuracy of $10^{-4}$ a.u. in 
the case of the CH$_2$ molecule calculated with 57 orbitals. 

From the technical point of view, one can start a DMRG calculation 
by setting $TRE_{\rm max}$ to $10^{-3}$ and when the algorithm has 
converged (the energy is unchanged, 
the number of block states are unchanged, the slope of the density 
matrix remains the same )
$TRE_{\rm max}$ can be adjusted by an order of magnitude until the desired
maximum value of the accuracy is reached. 
Using the calculated energy values and 
the truncation error obtained for various values of $TRE_{\rm max}$ 
(which is slightly below $TRE_{\rm max}$) 
the FCI energy can be estimated by Eq.~(\ref{eq:fcifit}). This equation
contains three free parameters ($E_{FCI}, a, b$) to determine from the fit.
However, based on our results we can set the parameter $a$ to one. 
We have found that one can gain one to three 
orders of magnitude improvements in the error of the correlation
energy by the
extrapolation method and 
that fixing the parameter $a$ to one always provides   
an upper bound. 
In order to obtain a more accurate fit one needs more data points, thus
$TRE_{\rm max}$ should
be adjusted in even smaller steps, especially, if the calculations are carried out only
up to a relative accuracy of $10^{-5}$, but we have not done such analysis yet. 

In case of solid state physics, chains with various lengths are 
calculated and the  
thermodynamic limit is extrapolated by the the so-called finite-size
scaling method. Using our procedure one can improve the energy values
obtained for a given length $L$ by two to three orders of magnitude, thus
the overall performance of finite-size scaling procedure can be 
improved significantly.

\subsection{Scaling of the number of block states}

As we have shown, the number of block states depends on the structure of the 
reduced density matrix spectrum. Thus it is not possible to determine the  
scaling behavior of the maximum number of block states as a function of the number of
orbitals and the fraction of filled orbitals in a rigorous way. On the other hand, 
in order to present a rough indication of computational resources used during
our calculations we have  
we have collected the values of the maximum number of block states selected dynamically 
by our the method in Table~\ref{block_states}. 

\subsection{Other factors that affects the accuracy} 

\begin{figure}
\includegraphics[scale=0.5]{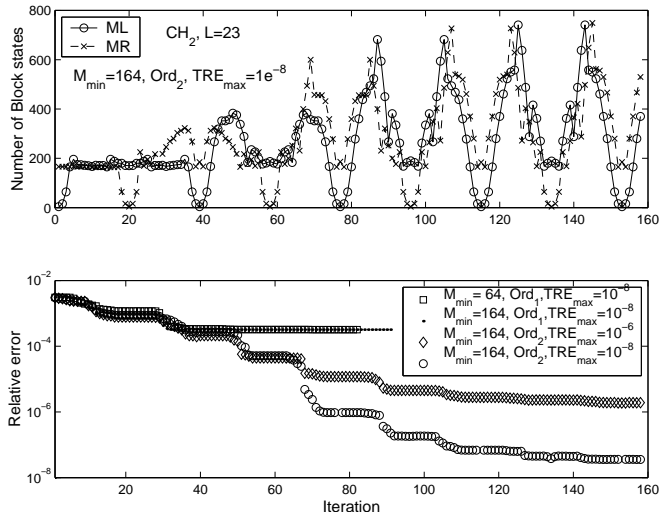}
\caption{The figure shows that an incorrect ordering can drive DMRG to a local minimum.}
\label{fig:ch2_23}
\end{figure}

It is important to note that our scaling results are obtained only for a
proper ordering of the orbitals in the initial chain.
We have found that for some cases  
the accuracy can be improved significantly if the HF levels were ordered with increasing
energy (labeled by $Ord_2$ on Fig.\ref{fig:ch2_23}) while for other cases we 
had to "mirror" the chains and placed orbitals occupied in the HF configuration 
to the center of the chain (labeled by $Ord_1$). A non-optimal ordering can in
fact lead the method 
to be trapped by a local minimum. This situation is shown explicitly   
in Fig.~\ref{fig:ch2_23} indicated by $Ord_1$. Even if $M_{\rm min}$ was almost tripled, 
the relative 
error converged to the same local minimum, which, on the other hand, also supports 
our previous statements that $M_{\rm min}$ does not affect the final convergence significantly. 
Changing the ordering, we have found for $Ord_2$ that the algorithm has always
converged to the value of $TRE_{\rm max}$.
Studying the optimal ordering can be a major field of research.   
Chan {\em et al.} \cite{chan} has 
already suggested a procedure to 
optimize the ordering in a recently published paper. We have not analyzed their solution yet.  

Another field of study can be the optimization of the superblock configuration.
In our largest calculations for 
the half-filled case (18 electron in 18 orbitals) the number of selected 
block states grew up to 1500--1800 with sizes of the Hilbert space of the 
superblock configuration increasing beyond 1.000.000. In order to decrease
the size of the Hilbert space we have derived an alternate protocol, modifying
the superblock configuration as 
$B_L \bullet B_R$ in a similar way as was done by Xiang. We found a considerably  
worse performance for these calculations. We believe that in the future it can worthwhile  
analyzing the speed of convergence for various superblock configurations.

\section{Summary}

We have applied the momentum-space version of DMRG in quantum chemistry in order to 
study the accuracy of the method. Analyzing the eigenvalue spectrum of the 
reduced density matrix and based on our previous results obtained for real 
space DMRG, we have shown that it is possible to set up the accuracy of the 
method in advance of the calculation by dynamically controlling the
truncation error and the number of block states. We have carried out
a detailed QC-DMRG study of the molecules 
H$_2$O, CH$_2$, and F$_2$ obtained with various basis sets   
in order to show that the relative error scales to the maximum threshold value
of the truncation error that was fixed in advance of the calculation. We
found that the 
linear relationship between the logarithm of the relative error and 
the logarithm of the maximum value of the truncation 
error is independent of the number of orbitals and the fraction of 
filled orbitals for the cases considered. Based on these results we 
have presented a novel approach to
extrapolate the FCI energy, a method which could also improve the accuracy of 
the finite-size scaling method when $k$-DMRG is applied in solid state physics. 
We have addressed new problems related to the inaccuracy of starting block
state configuration and presented solutions
to achieve faster convergence and better stability of the target state. 

The maximum number of block states that the 
algorithm selected out in the dynamic fashion 
was in the range of 1500--2000, the largest size of the  
Hilbert space related to the superblock Hamiltonian was 800.000--1.200.000, 
and the longest chains that we have studied contained 57 sites.

Although momentum is not a good quantum number if $k$-DMRG is applied in 
quantum chemistry, there are still a few remarks which might indicate why 
QC-DMRG method can work well in the field: 

$\bullet$ In most of the cases the calculations are carried out in the small 
$U$ limit known to converge fast.
  
$\bullet$ The number of electrons is fixed for a given molecule. Therefore, 
doubling the length of the system will not imply in general keeping the
fraction of filled orbitals fixed. 
Thus calculating a molecule with more basis states would mean 
longer chains but with lower filling value which usually has a better 
convergence.

$\bullet$ The practical use of DMRG in quantum chemistry
can open a route to active spaces well beyond today's limits, yielding 
complete active space self-consistent field (CASSCF) solutions with a relative
error of the correlation energy of the order of $10^{-4}$ to $10^{-5}$. 
This can be realized by a few thousand block states, which also expected to 
hold for longer chains as well. Therefore, we believe that 3000--4000 block 
state will provide satisfactory results for all the chain lengths and fillings
which are of interest in the immediate future.
 
$\bullet$ Although the structure of the Hamiltonian is very complicated, it 
is decomposed into several parts. This means that during the diagonalization 
step each component of the Hamiltonian can be applied on the wavefunction 
independently, therefore, the method is an excellent candidate for parallel 
computers.

Our source code was written in the the framework of the {\em Matlab} 
programming environment
and the C++ code as well as the standalone code was produced by the 
Matlab compiler. Most of our numerical calculations were carried out 
on Athlon XP 1800+ processors under Linux and in some cases on a SGI 3000
machine of the local computer center. 
For the largest calculations comprising
$M$=1700--2000 block states (F$_2$ 18/18) the program required 200--500 MB of RAM, running about
50--60 hours on Athlon XP 1800+ processor to achieve $10^{-4}$-$10^{-5}$ a.u.
absolute accuracy. The scaling of computational time with the number of orbitals
still can not be determined because of the development stage of our code,
but as a rough indication it took some 15 hours for 8/24 chain and more than a 
week for the 6/57 case. 
The present stage of our code limited the number of block states 
around 2000, however, solving a few technical points we expect that the feasible $M$  
can be increased significantly in the future.

\acknowledgements

This research was supported in part by the Fonds der Chemischen Industrie and  
the Hungarian Research Fund(OTKA) Grant No.\ 30173 and 32231. 
\"O.~L.~ also acknowledges the useful discussions with J.\ S\'olyom and G.\ F\'ath. 

\bibliography{qcdmrg12}
\begin{table}
\begin{tabular}{lcccccrr}
&
basis set &
bond distance &
bond angle &
electrons &
orbitals &
\multicolumn{1}{c}{HF energy} &
\multicolumn{1}{c}{FCI energy} \\
&reference& (a.u.) &
   \multicolumn{1}{c}{(degrees)} &&&
   \multicolumn{1}{c}{(a.u.)} &
   \multicolumn{1}{c}{(a.u.)} \\[8pt]
H$_2$O & Double-Zeta\cite{dzw1} & 1.84345 & 110.565 & 10 & 14 & $-76.009838$ & $-76.157866$ \\
H$_2$O & DZP\cite{Bauschlicher-Taylor-1986} & 1.88973 & 104.500 & 8 & 25 (24) & $-76.040551$     & $-76.256634$   \\
 CH$_2$ $^1A_1$ & DZ\cite{Huzinaga,Dunning}    & 2.02230 & 129.4667 & 6   & 14 (13)   & $-38.909437$ & $-38.932107$ \\
 CH$_2$ $^3B_1$ &             &        &          & 6   & 14 (13)  &     & $-38.979393$ \\
 CH$_2$ $^1A_1$ & cc-pVDZ\cite{cc-pVTZ}& 2.02230   & 129.4667 & 6   & 24 (23)  & $-38.921647$ & $-39.006652$ \\
 CH$_2$ $^3B_1$ &             &        &          & 6   & 24 (23)  &     & $-39.041774$ \\
 CH$_2$ $^1A_1$ &  cc-pVTZ\cite{cc-pVTZ} & 2.02230 & 129.4667 & 6 & 58 (57)  & $-38.932575$ & $-39.087006$ \\
 F$_2$          & DZ\cite{Huzinaga,Dunning} & $2.64373$ &        & 14  & 20 (14)  & $-198.707822$     & $-198.915252$        \\
 F$_2$    & split valence\cite{Ahlrichs-SV} & $2.68797$ &        & 18  & 18  & $-198.484167$ & $-198.761551$ \\
\end{tabular}
\vskip 8pt
\caption{
Geometries and benchmark energy values for the calculated
molecules. The number of correlated orbitals
is given in parentheses, unless it agrees with the total number of orbitals.
}
\label{molecules}
\end{table}                    

\begin{table}
\begin{tabular}{lrrrrrrr}
                 &        &        &       &         &         &       &       \\
                 & CH$_2$ & H$_2$O & F$_2$ & CH$_2$  & H$_2$O  & F$_2$ & CH$_2$\\ 
L                & 6/13   &  10/14 & 14/14 & 6/23    & 8/24    & 18/18 & 6/57  \\
Filling          & 0.230  & 0.357  & 0.500 & 0.130   & 0.166   & 0.500 & 0.052 \\\hline
$\Delta E_{Abs}$ & $M_{\rm max}$ & $M_{\rm max}$ & $M_{\rm max}$ & $M_{\rm max}$ & $M_{\rm max}$ & $M_{\rm max}$ & $M_{\rm max}$\\\hline 
$10^{-2}$        &   25   &   40   &  280  & 150     &  130    &  170  & 300   \\
$10^{-3}$        &   40   &   60   &  350  & 280     &  320    &  520  & 480   \\
$10^{-4}$        &  100   &  140   &  800  & 370     &  440    & 1100  & 620   \\
$10^{-5}$        &  160   &  300   & 1450  & 580     &  650    & 1800  &       \\
$10^{-6}$        &  230   &  420   &       & 670     &  820    &       &       \\
$10^{-7}$        &  300   &  530   &       & 720     &         &       &       \\
$10^{-8}$        &  360   &  650   &       & 880     &         &       &       \\
$10^{-9}$        &  420   &  780   &       &         &         &       &       \\
\end{tabular}
\vskip .1 truein 
\caption{The maximum number of the block states selected out dynamically by DMRG 
to reach a given value of absolute accuracy.   
The second row contains the number of electrons and orbitals of 
each test calculations and below the fraction of filled orbitals is listed.}
\label{block_states}
\end{table}

\end{document}